\documentclass[sigconf]{acmart}
\AtBeginDocument{%
  \providecommand\BibTeX{{%
    \normalfont B\kern-0.5em{\scshape i\kern-0.25em b}\kern-0.8em\TeX}}}
    
\acmConference[]{}{}{}

\acmDOI{}

\acmISBN{}

\acmPrice{}

\usepackage{amsmath,amssymb,amsfonts}
\usepackage{algorithmic}
\usepackage{graphicx}
\usepackage{textcomp}
\usepackage{xcolor}
\usepackage{multirow}
\usepackage{url}
\usepackage{times}
\begin{document}

\title{AI on the Edge: Rethinking AI-based IoT Applications Using Specialized Edge Architectures
}

\author{Qianlin Liang}
\affiliation{\institution{University of Massachusetts, Amherst}}
\email{qliang@cs.umass.edu}

\author{Prashant Shenoy}
\affiliation{\institution{University of Massachusetts, Amherst}}
\email{shenoy@cs.umass.edu}

\author{David Irwin}
\affiliation{\institution{University of Massachusetts, Amherst}}
\email{irwin@ecs.umass.edu}

%

\begin{abstract}
Edge computing has emerged as a popular paradigm for supporting mobile and IoT applications with low latency or high bandwidth needs. The attractiveness of edge computing has been further enhanced due to the recent availability of special-purpose hardware to accelerate specific compute tasks, such as deep learning inference, on edge nodes. In this paper, we experimentally compare the benefits and limitations of using specialized edge systems, built using edge accelerators, to more traditional forms of edge and cloud computing. 
Our experimental study using edge-based AI workloads shows that today's edge accelerators can provide comparable, and in many cases better, performance, when normalized for power or cost, than traditional edge and cloud servers. They also provide latency and bandwidth benefits for split processing, across and within tiers, when using model compression or model splitting, but require dynamic methods to determine the optimal split across tiers. We find that edge accelerators can support varying degrees of concurrency for multi-tenant inference applications, but lack isolation mechanisms necessary for edge cloud multi-tenant hosting. 
\end{abstract}

\maketitle



\section{Introduction}

\begin{table*}[ht]
\centering 
   \begin{tabular}{|l | l l l l|} \hline  
        Device & Power (W) & Memory   & Cost & Accelerated  Workloads  \\ \hline
        Intel NCS2 VPU & 1 - 2 & 512 MB  & \$99 & vision, imaging  \\
        Google EdgeTPU & 0.5 - 2 & 8MB & \$75 & any TensorFlow lite model  \\
        Nvidia  Nano &  5 - 10 & 4 GB & \$99 & any GPU workload; AI \\
        Nvidia  TX2 &  7.5 & 8 GB & \$399 & any GPU workload; AI \\ \hline
   \end{tabular}
   \caption{Characteristics of edge accelerators}
   \label{tab:edge_accelerators_profiles}
\end{table*}

Edge computing has recently emerged as a complement to cloud computing for running online applications with low latency or high bandwidth needs~\cite{emergence-of-edge-computing}. Internet of Things (IoT) and mobile applications are particularly well-suited for the edge computing paradigm, since they often produce streaming data that requires real-time analysis and control, which can be optimally performed at the edge.  

Conventional edge computing comes in many different flavors.
Cloudlets~\cite{cmu-cloudlet-paper} represent one popular paradigm of edge computing that entails deploying server clusters at the end-points of the network; by deploying traditional servers at the edge, cloudlets enable ``server-class'' applications to be deployed at the edge rather than the cloud.   Edge gateways represent a different flavor of edge computing that involves deploying embedded nodes, individually or in groups, to serve as the hub for 
applications such as smart homes. Such edge gateways provide more limited compute capabilities at the edge,  but nevertheless provide useful functionality, such as data aggregations and local on-node processing for certain low-latency tasks. 

These two flavors of edge computing offer very different tradeoffs. The latter paradigm utilizes small form-factor hardware (e.g., Raspberry Pi-class nodes), has low cost, low power consumption and also constrained compute capabilities, which increases reliance on the cloud. Cloudlet-style edge computing, on the other hand, provides much greater compute capabilities at the edge, but incurs higher hardware costs,
larger form factor servers, and higher power consumption; there is also less reliance on the cloud for many applications. 

\begin{figure}[t]
  \centering
   \includegraphics[width=2in]{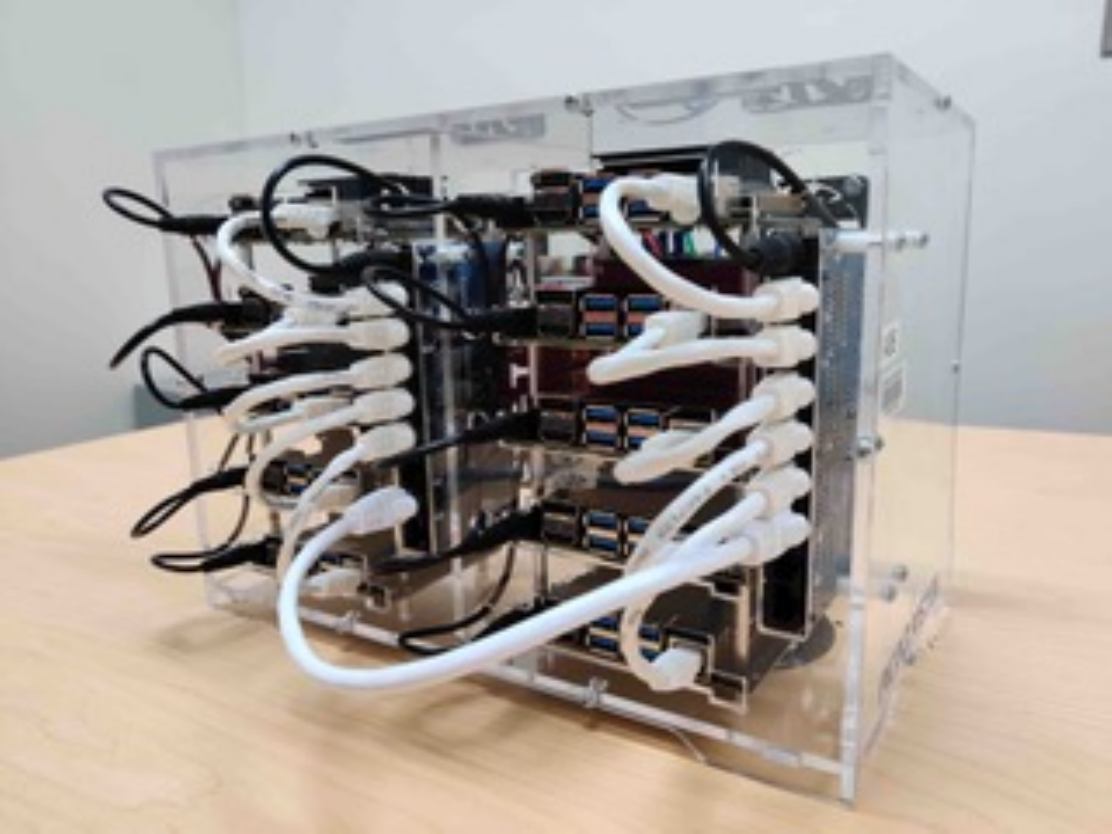}
   \caption{A 10-node cluster of low-power Jetson nano GPUs.}
   \label{fig:nano}
 \end{figure}

Recently a third flavor of edge computing has emerged that combines the key advantages 
of both paradigms. This paradigm, which we refer to as specialized edge architectures, 
has become possible with the advent of special-purpose hardware designed to accelerate specific compute- or I/O-intensive operations.  In particular, a number of edge
hardware accelerators, such as Intel's Movidius Vision Processing Unit (VPU)~\cite{movidius_vpu}, Google's 
Edge Tensor processing Unit (TPU)~\cite{tpu}, Nvdia's Jetson Nano and TX2 edge GPUs~\cite{nano,tx2}, and Apple's Neural  Engine have emerged recently. These accelerators are explicitly designed, and marketed by their vendors, for edge computing, with the specific goal of supporting {\em edge-based AI} applications such as computer vision, visual and speech analytics, and deep learning inference. 
By customizing silicon to a single, or a small, class of applications, these hardware accelerators claim to provide major performance improvements at much lower cost and energy points when compared to traditional general-purpose hardware. As a result, it is now possible to embed ``wimpy'' edge nodes with these accelerators and approach the compute 
capabilities of general-purpose servers (e.g., cloudlets) for specific applications.\footnote{Of course, cloudlets can also be equipped with hardware accelerators, further enhancing their capabilities.} Figure~\ref{fig:nano} depicts a 10 node cluster of low-end Pi-class nodes equipped with Jetson Nano GPUs; this entire embedded GPU cluster costs about \$1,500 (or approximately the cost of a single traditional server), consumes only 90w at full GPU load,  and measures 13x8x8 inches, an order of magnitude smaller footprint than a server rack.  As a result, it opens up new possibilities for edge deployments in power-constrained or space-constrained settings that are not feasible with  conventional flavors of edge computing.

In this paper, we address the question of how to rethink the design of edge-based AI applications in light of specialized edge architectures? Using an empirical approach, we seek to quantitatively understand the benefits and limitations of these architectures when compared to more traditional edge and cloud-based systems. 
In particular, we seek to answer three sets of research questions: (1) What are the price, performance, and energy tradeoffs offered by emerging edge hardware accelerators when compared to traditional edge and cloud computing? (2) How should modern IoT applications exploit the distributed processing capabilities of specialized edge nodes and the cloud  by
employing various types of split processing? (3) How suitable are edge accelerators for supporting concurrent edge applications from multiple tenants?

We seek to answer these questions through the lens of a particular class of applications---edge-based vision and speech processing---using  an experimental testbed of several different edge accelerators and embedded nodes. Our results show that edge accelerators can yield up to 10-100$\times$ better normalized performance, on a performance-per-watt and performance-per-dollar basis, than general-purpose edge servers. They also show that split processing on machine learning inference, using  model compression and model splitting,
 between device-edge, edge-edge, and edge-cloud tiers
can yield significant bandwidth savings and latency benefits. Since the benefit can vary by the model and workload, we also find that such split processing must be done carefully on a per-application basis  to  maximize benefits. Finally,
we find that systems optimizations such as model quantization and RAM model swapping can enhance the degree of concurrency supported by edge accelerators but that their 
lack of performance isolation and security can be a hurdle. Overall, our results show the significant promise for specialized edge architectures, but
also point to the need to address open research  questions to fully realize their potential.

\section{Background}
\label{sec:background}

In this section, we present background on cloud- and edge-based IoT applications as well
as specialized edge architectures for edge-based AI applications.

\begin{figure*}[ht]
    \centering
    \begin{tabular}{cccc}
    \includegraphics[width=0.55in]{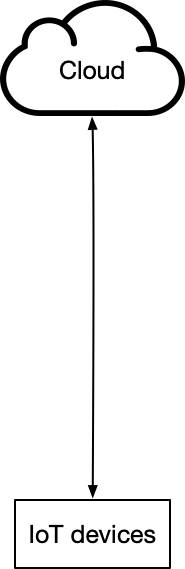} & 
     \includegraphics[width=0.55in]{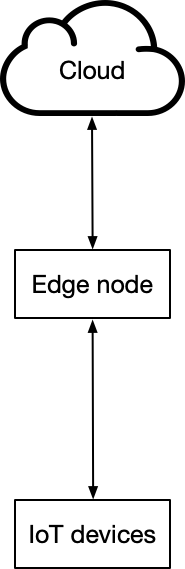} &
    \includegraphics[width=0.55in]{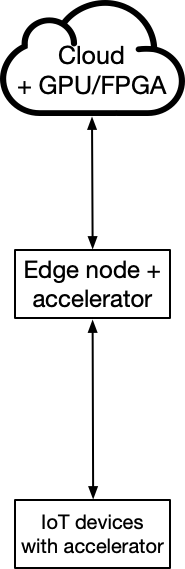} & 
  \includegraphics[width=0.6in]{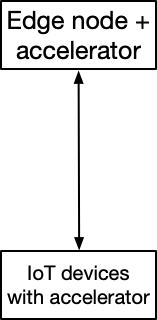}\\
  (a) Two tier \hspace*{0.3in} & (b) Three tier \hspace*{0.2in} & (c) Specialized three-tier \hspace*{0.2in} & (d) Specialized two-tier
  \end{tabular}
    \caption{Tiered architectures for IoT applications that use the device, edge, and cloud.}
    \label{fig:architectures}
\end{figure*}

\noindent\textbf{Cloud- and Edge-based IoT Applications:}
Many IoT devices with networking (e.g., WiFi) capabilities  employ a {\em two-tier} cloud architecture depicted in Figure \ref{fig:architectures}(a), where the device transmits data to the cloud for processing. Examples of such IoT devices include the Nest thermostat \cite{Nest}, Wemo smart switch, and LiFX smart lights.  It is also increasingly common for 
IoT devices to  use a {\em three-tier} architecture, depicted in Figure \ref{fig:architectures}(b), that leverages both the edge and the cloud\cite{hung2018videoedge}. 
Application  processing is split between the edge and the cloud, with the edge performing some initial processing of the data and the cloud providing more substantive processing capabilities.  Battery-powered IoT devices, such as smart door locks that use low-power wireless protocols (e.g.,  Bluetooth LE), employ a three tier architecture and rely on an intermediate edge node~\cite{smarthings-hub,wink-hub,Mortazavi:2017:CMC:3132211.3134464} to provide a gateway to the cloud.  Edge computing has also shown promise for applications, such as augmented and virtual reality (AR-VR)~\cite{satya-cmu} \cite{Zhang:2017:TEE:3132211.3134463}, computation offloading~\cite{Chandra-offload,Cuervo:2010:MMS:1814433.1814441} \cite{Dong:2017:EOF:3132211.3134448}, and online gaming \cite{Google-stadia-gaming}, which use Cloudlet-style edge clusters with more substantial compute capabilities to provide low latency processing.

\noindent \textbf{Edge-based AI workloads:} 
An emerging class of edge workloads, referred to as ``AI on the Edge'' or edge-based AI, involves running machine  learning or deep learning inference on edge nodes. Some researchers have argued that such visual analytics and machine learning inference on edge nodes is poised to become the ``killer app'' for edge computing \cite{ananthanarayanan2019demo}\cite{ananthanarayanan2017real-time}. 
This application use case has become promising due to the proliferation of smart cameras and smart voice assistants that generate significant amounts of video and audio data, which requires  vision and speech processing in real time.  Doing so involves deploying previously-trained  deep learning models at the edge to perform near real-time inference or predictions on the video and audio data. Such inference may involve tasks, such as  image classification or object detection in  video feeds~\cite{bandwidth-efficient-video-analytics-edge-computing,lavea,grassi2017parkmaster,Kar:2017:RTE:3132211.3134461} or speech recognition from voice assistants to understand spoken commands---all of which have low-latency and near real-time constraints.\footnote{For example, a user who uses a voice assistant to turn on a smart light bulb using a spoken command expects the lights to turn on in near real time. Similarly, smart cameras send real-time push notifications when they detect something suspicious in their video feed, which requires low-latency real-time processing of video.} 

\noindent \textbf{Special-purpose edge computing and edge  accelerators:} 
Specialized edge computing has emerged as a new paradigm in edge computing with the advent of edge accelerators that target acceleration of machine learning and  deep learning inference tasks. Figures \ref{fig:architectures}(c) and (d) depict edge computing with specialized architectures, where one or more tiers (device, edge, cloud) employ hardware accelerators. Each tier can leverage such specialized hardware, when available, to either boost the processing  capabilities of that tier, which implies that each tier has less reliance on higher-level tiers. Figure~\ref{fig:architectures}(d) is a special case of Figure~\ref{fig:architectures}(c), where all application processing is performed on the device  or on the edge using specialized hardware. In scenarios where the specialized edge is a 
cluster, as in Figure~\ref{fig:nano}, more than one edge node may be leveraged for distributed edge processing. 

Table \ref{tab:edge_accelerators_profiles} list various edge accelerators and their characteristics. Intel's Movidius Neural Compute Stick (NCS) employs a Vision Processing Unit (VPU) to accelerate deep learning models for computer vision tasks, such as object detection and recognition~\cite{movidius_vpu}. Google's Edge Tensor Processing Unit (TPU)~\cite{tpu} can accelerate any Tensorflow ML model inference as long as it is compatible with the Tensorflow-lite framework. Nvidia's edge GPUs include the Jetson Nano GPU~\cite{nano}, as well as the Jetson TX2~\cite{tx2} GPU, which are both designed to provide full-fledged GPU capabilities on low-end edge nodes with a smaller power footprint than desktop- and server-class GPUs. From a power standpoint, Nvidia's Jetson Nano uses a default power budget of only 5W, which is up to 40$\times$ lower than desktop-class GPUs, while Google's TPU uses a power budget of only 2W. From a performance standpoint, all of these hardware accelerators promise large performance improvements
for low-end edge nodes and, in some cases, server-like performance, even when running on low-end Raspberry PI-class nodes. 
Specialized hardware is also becoming available for end-devices, which allows the processing to be done on the device itself, when appropriate, rather than sending data to edge or cloud servers. Examples include the Sparkfun Tensorflow-lite hardware board for micro-controller-based IoT devices~\cite{sparkfun} and the GAP8 IoT processor~\cite{GAP8}


\noindent \textbf{Split processing:}  Edge architectures have been employed for various forms of distributed processing, with application processing split within and across tiers. Processing may be split across
device, edge and  cloud tiers by leveraging specialized hardware at each tier,
yielding {\em vertical} splitting.
 Processing at each tier can be further split across nodes within that tier to leverage multiple hardware accelerators, yielding {\em horizontal} splitting.  Model compression \cite{Teerapittayanon2017DistributedDN} and model splitting \cite{Kang:2017:NCI:3093315.3037698}  are examples of distributed ML inference that use such split processing.

\begin{table*}[t]
    \centering
    \begin{tabular}{l| l l l l l l} \hline
         Workload & Model  &  Input size & Model  & Params  & \# Float operations & Depth\\ 
                & name &     & size (MB)  & (M) & per inference (M) & multiplier \\
         \hline
     \multirow{2}{*}{Image Classification} &  MobileNet V2 &  $224\times224\times3$ & 14 & 3.54 & 602.29 & 1.0 \\
         & Inception V4 &  $299\times299\times3$ & 163 & 42.74 & 24553.87 & - \\ \hline
     \multirow{2}{*}{Object Detection} &  SSD MobileNet V1  & $300\times300\times3$ & 28 & 6.86 & 2475.24 & 1.0 \\
         & SSD MobileNet V2 & $300\times300\times3$ & 66 & 16.89 & 3751.52 & 1.0 \\ \hline
     Keyword Spotting &  cnn-trad-fpool3  & $99\times40$ & 3.6 & 0.94 & 410.89 & - \\ \hline
    \end{tabular}
    \caption{Characteristics of the deep learning models used in our study.}
    \label{tab:model_profile}
\end{table*}


\section{Experimental Methodology}


\noindent \textbf{Problem statement:} The goal of our work is to empirically study the feasibility of using a hardware-accelerated specialized edge tier to provide ``server-class'' performance of cloudlet-style edge servers at the cost, power, and form-factor of  Pi-class edge nodes, with a specific emphasis on edge-based AI workloads.  To do so, our study addresses the following  questions: (1) What are the price, performance, and energy
benefits, if any, offered by  edge hardware accelerators when compared to general-purpose edge and cloud computing? How do specialized edge nodes compare to traditional edge nodes
with respect to raw performance and normalized performance-per-watt and performance-per-dollar? How do these benefits vary with different workloads, such as
image/video and audio processing, and different deep learning models?  (2) How should
IoT application exploit distributed and split processing capabilities offered at various tiers? How are the benefits and overheads of splitting application processing over 
centralized processing at a single tier? Are there scenarios where performing data processing at a single tier is better than splitting application processing across tiers?
(3) How capable are these edge accelerators for supporting concurrent model execution to provide multi-tenancy in edge clusters?   

\noindent \textbf{Experimental setup:} Our experimental setup comprises a small cluster of 
single-board computing (``Pi-class'') nodes that are equipped with 
  four  edge accelerator platforms:  Intel Movidius NCS2 VPU, Google Edge TPU, Nvdia Jetson Nano GPU, and Nvidia TX2 GPU. To compare with more traditional edge architectures, we also consider a Raspberry Pi3 node as an example of a resource-constrained edge device, and an x86 server with a 3.0GHz Xeon Skylake CPU as an example of a cloudlet-style edge server. We also consider a NVIDIA Tesla V100 GPU on  Amazon EC2  \texttt{p3.2xlarge} cloud instance to mimic a specialized edge server or specialized cloud server.

\noindent \textbf{Workloads:} Our workload  consists of three common vision-based image processing and speech-based audio-processing tasks that arise in many edge-based AI applications:
\begin{itemize}
\item {\em Image classification:} The goal of image classification is to assign a text label (i.e., ``classify'') to an image based on its contents. For example, a label such as
``apple",  ``dog" or ``car" may be assigned by the classifier based on the image. Typically model inference yields multiple labels with probabilities on the likely contents of the image. 

\item {\em Object detection:} Object detection is a harder task than classification since it involves determining all objects of interest that are present in the image, by computing a bounding box around each such object, and then  assigning a probabilistic  label to each object.

\item {\em Keyword spotting:} Keyword spotting involves 
processing an audio stream to detect and recognize the occurrence of a set of keywords (e.g., "Hey Siri" function on iPhone).
\end{itemize}
All three workloads use deep learning models, and there has been a wealth of research 
on these problems over the past decade \cite{DeepLearningTextBook}. Pre-trained deep learning models are now available 
for these tasks from multiple sources and these models are designed to run on a variety
of hardware and software platforms.  We use these pre-trained models for our micro-bechmarking study since it allows us to run the same standard model on all hardware devices, and also enables others to repeat our experiments. Our experiments use the following 5 models:  MobileNet V2 and Inception V4 for image classification, 
SSD MobileNet V1 and SSD MobileNet V2 for object detection, and \texttt{cnn-trad-fpool3} in \cite{google_keyword_spotting} for keyword spotting. Table \ref{tab:model_profile} lists the key characteristics of the models along with the default model configurations  used in our experiments.   


\section{Performance and Energy Microbenchmarks}
\label{sec:micro}

\begin{figure*}[t]
    \centering
    \begin{tabular}{c c c}
         \includegraphics[width=2.25in]{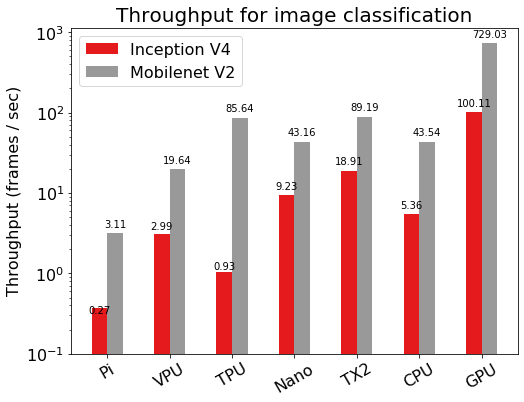} &
         \includegraphics[width=2.25in]{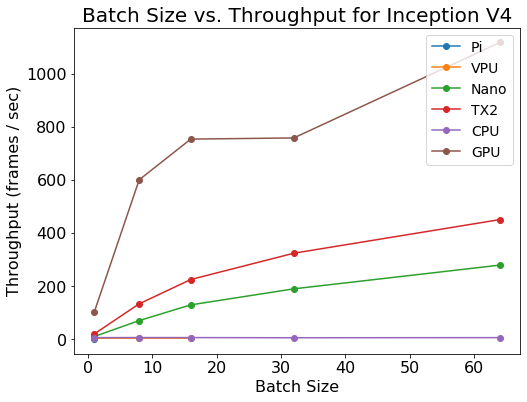} & \includegraphics[width=2.25in]{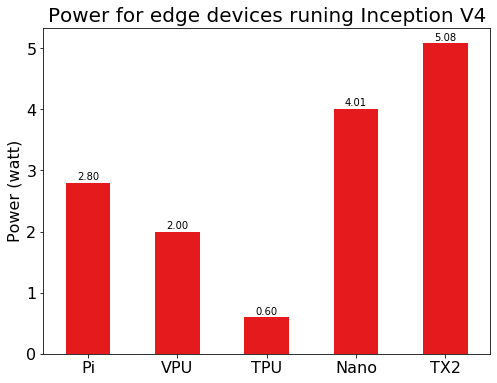} \\
         (a) Throughput & (b) Batch Throughput & (c) Power consumption 
    \end{tabular}
    \caption{(a)Throughput of edge and cloud devices for image classification. 
    (b) The impact of batch size for edge and cloud devices for the Inception V4 model (c) Power consumption of edge devices for the Inception V4 model. The server CPU and GPU consume 131.26W and 111.66W, respectively, for the same model.}
    \label{fig:throughput_batch_power}
\end{figure*}

\begin{figure*}[t]
    \centering
    \begin{tabular}{c c}
         \includegraphics[width=2.75in]{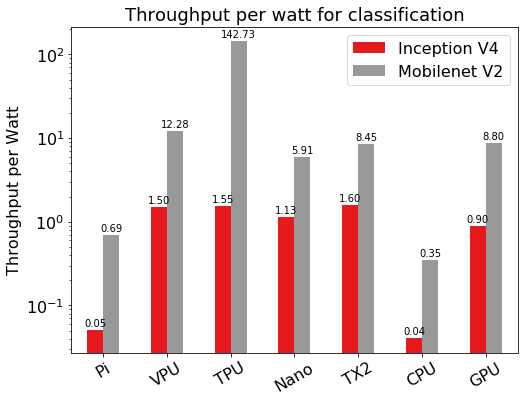} &
         \includegraphics[width=2.75in]{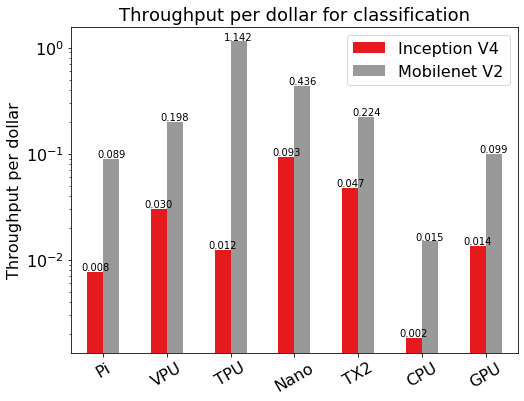} \\ 
         (a) Throughput per watt & (b) Throughput per unit cost 
    \end{tabular}    
    \caption{Normalized performance per watt and per unit cost for various devices. Unit cost data is shown in Table \ref{tab:edge_accelerators_profiles}}
    \label{fig:normalized_throughput}
\end{figure*}

Our first experiment involves comparing raw and normalized performance and power of 
specialized edge nodes to more traditional edge architectures comprising (i) resource-constrained edge nodes (Pi3), (ii) x86 server-based edge nodes (``cloudlet server''),
and (iii) GPU-equipped x86 servers.  We microbenchmark various edge nodes under
our three workloads (classification, object detection and keyword spotting) and the 
corresponding models shown in Table \ref{tab:model_profile} and measure
 throughput and power consumption under these workloads. 

\noindent \textbf{Methodology:} To ensure a fair comparison across hardware platforms, we run the \emph{same} model on all platforms and subject it to the \emph{same} inference workload. For object classification
and object detection, we use the CAVIAR test case Scenarios dataset \cite{CAVIAR} as our 
inference workload. For keyword spotting, we use the Speech Commands dataset \cite{DBLP:journals/corr/abs-1812-01739} as our inference workload.  Although the model
and the inference workload used to drive the model are identical on all platforms, it should be noted that the deep learning (DL) software platform used to execute this model varies by device.
This is because there is no single  DL software platform that runs well on all hardware accelerators. While TensorFlow runs on many of our devices, we found that it almost always had worse performance than the native vendor-designed tool for running DL inference. 

Thus, we choose the native vendor-recommended software DL platform for each device since it yields the maximum throughput and best results. 
Specifically, we use Intel Openvino \cite{Openvino} for the Intel VPU,
the specialized \emph{edgetpu} software module for Google's Edge  TPU, and TensorRT  \cite{tensor-rt}
for Nvidia's Jetson Nano, TX2 and cloud GPUs. Finally, we use TensorFlow to execute our
models on all CPUs, namely Raspberry Pi3 and Intel Xeon CPU.  Our throughput microbenchmark, written in python, iteratively involves making inferences using the above inference workloads and computes throughput in term of inferences per second. 
In addition to measuring sequential inference throughout, 
we also measure the impact of batching inference requests on the throughput---since batching is often used in production settings to enhance the throughput of deep learning model inference. Our power microbenchmarks measure the mean power consumption as well
as the total energy consumed during an individual inference request.

 We use a combination of hardware and software tools for our power microbenchmarks. For USB devices such as Intel VPU and 
Google EdgeTPU, we use a USB power meter with data logging capabilities to measure 
the energy used and instantaneous power consumption during inference. For NVidia GPUs, 
we use \texttt{nvidia-smi} software profiling tool that provides power statistics for NVidia GPUs \cite{nvidia-smi}. For the 
cloud-based Intel Xeon CPU and Raspberry Pi CPU, we use the Turbostat Linux profiling tools \cite{Linux-turbostat} to measure the CPU power usage; Turbostat also works in virtualized environments such as cloud servers for power profiling.



\noindent \textbf{Performance results:}
We begin with microbenchmarking our hardware accelerators using the image classification workload. Figure \ref{fig:throughput_batch_power} shows the throughput and power usage results for our two image classification models: Mobilenet V2 and Inception V4. As shown in Table \ref{tab:model_profile}, Inception 
is a more complex model that is around 7$\times$ larger in size and parameters than Mobilenet.
Figure~\ref{fig:throughput_batch_power}(a) depicts the mean inference throughput in terms of frames/s for various hardware accelerators running these models; note the log scale on the y-axis depicting 
throughput. 

The figure yields the following observation: (1)  All four edge accelerators
provide a significant increase in performance when compared to a vanilla Pi3 edge node, yielding between 6$\times$ to 28$\times$ throughput increase for Mobilenet and 3.4$\times$ to 70$\times$ throughput increase for Inception. (2) Interestingly, some of the edge accelerators even outperform a modern Xeon processor, which align with their claims of ``server-class'' performance using low-cost hardware. Both Nvidia GPUs outperform the x86 CPU by 1.7$\times$ to 3.5$\times$
for MobileNet and have comparable to 2$\times$ higher throughput for Inception.  The VPU is the slowest of the four and yields about half the CPU throughput, while the TPU is 5$\times$ slower for Inception but 1.9$\times$ faster for Mobilenet. (3)  Not surprisingly, the cloud GPU still holds a significant performance advantage over all edge accelerators with 5$\times$ to 8$\times$ 
 higher throughput than the fastest edge accelerator (TX2). 
 
 While the throughput microbenchmarks above assume sequential inference requests, we next measure throughput using input batching. Batching of multiple inputs enables the hardware accelerator to parallelize model inference, thereby increasing hardware utilization and
 the resulting throughput. We vary the input batch size from 1 to 64 and measure the inference throughout for different hardware accelerators. Figure~\ref{fig:throughput_batch_power}(b) depicts the throughput results for the Inception model (results for Mobilenet are similar and omitted due to space constraints). The figure shows that batching is very effective for all GPUs; the throughout increases with batch size but shows diminishing improvements beyond a batch size of 16.  A batch size of 16 yields 13.94$\times$  and 11.85$\times$ throughput increase for Jetson Nano and TX2 GPUs, while a batch size of 64 yields 30.17$\times$ and 23.79$\times$.
 We also find that batching is not  effective for the TPU or VPU. We attribute this behavior to the smaller memory capacity of these devices that reduces their effectiveness 
 for batched input processing.
 
\noindent \textbf{Power results:}
Finally, Figure~\ref{fig:throughput_batch_power}(c) plots the mean power consumption of various hardware devices when performing 
inference.\footnote{The plot depicts power consumption of only the accelerator or the CPU and does not include power consumed by the rest of the node or its peripherals.} As noted earlier, a combination of USB power meters and Linux and NVidia profiling tools were used to 
measure power usage.  We also measured the total energy consumed per inference request but omit those results here since they directly correlate to the mean power  usage.
As shown in the figure, the TPU is the most power-efficienct
device and consumes only $0.6$ watts during inference, with the VPU being the next most
power efficient with a power consumption of $2$ watts. The Jetson Nano and TX2 GPUs
consume $4.01$ and $5.08$ watts on average during inference. In contrast, the Tesla Cloud GPU  and the Intel Xeon CPU consume $111.66$ and $131.26$ watts during inference, significantly higher than the edge accelerators.

\begin{table*}[t]
    \centering
    \begin{tabular}{l | l l l l l l l l} \hline
         Workloads & Models & Pi & VPU & TPU & Nano & TX2 & CPU & GPU \\ \hline
         \multirow{2}{*}{Image Classification} & MobileNet V2 & 3.11 & 19.64 & 85.64 &	43.16 &	89.19 &	43.54 &	729.03 \\
         & Inception V4 & 0.27 &	2.99 &	0.93 &	9.23 &	18.91 &	5.36 &	100.11 \\ \hline
         \multirow{2}{*}{Object Detection} & SSD MobileNet V1 & 1.39 &	10.66 &	21.09 &	23.97 &	46.90 &	21.23 &	499.83 \\
         & SSD MobileNet V2 & 1.10 &	8.37 &	17.90 &	19.64 &	36.34 &	17.44 &	372.74\\ \hline
         Keyword Spotting & cnn-trad-fpool3 & 15.85 &	26.65 &	33.31 &	299.98 &	449.15 &	201.33 &	2314.91\\ \hline
         
    \end{tabular}
    \caption{Throughput in inferences per second}
    \label{tab:throughput}
\end{table*}

\noindent \textbf{Normalized performance:}
Figure \ref{fig:normalized_throughput}(a) and (b) plot the normalized throughput of various hardware devices with respect to power and cost. The normalized metrics of performance per watt and performance 
per dollar, respectively, enable a different comparison of these devices in constrast  to using raw performance or power. Figure \ref{fig:normalized_throughput}(a) plots the throughput per watt for various devices. When normalized for power consumption, {\em all edge accelerators outperform the x86 CPU by 10-100$\times$ and  become comparable
or  outperform the cloud GPU}. Due to their low power consumption, the TPU and VPU offer the highest performance per watt across all devices. Overall, the performance per watt is  25.5 to 77\% higher for the various edge accelerators when compared to the cloud GPU for the Inception workload. For Mobilenet, the TPU and VPU 
yield a 16$\times$ and 1.3$\times$ better performance per watt than the cloud GPU, respectively. 
Figure \ref{fig:normalized_throughput}(b) plots the throughput per dollar cost for all devices. Once again, we see that
all edge accelerators provide a higher  throughput per dollar cost  than the cloud GPU and x86 CPU due to their low cost.   Even the TX2 GPU, which has  a relatively high list price of \$399,  yields a 1.3$\times$ better performance per dollar cost than the cloud GPU.

Next, we repeat the above experiments for the object detection and keyword spotting workloads. Table \ref{tab:throughput} summarizes the inference throughput obtained for various hardware devices under various deep learning models and workloads. While there are some variations 
in throughput across audio and image workload and different models, the broad results from
Figure \ref{fig:throughput_batch_power} hold for these results. All edge accelerators provide very significant throughput  improvements over low-end edge nodes, such as the Raspberry Pi, and many outperform
even a x86 server processor. Broadly, the TX2 edge GPU provides the highest throughput 
across the four edge devices; performance can be  
roughly ordered as VPU, TPU, Jetson Nano, and TX2  for various workloads. 
The cloud GPU continues to provide the greater raw performance across all devices, but becomes comparable or slightly worse than the accelerators on a 
a normalized performance per watt and performance per dollar basis (not shown here to due to space constraints)---similar to the trends 
shown in Figure \ref{fig:normalized_throughput}.

{\noindent \bf Key takeaways:} On a raw performance basis, we see a rough performance 
order across edge accelerators for inference workloads, namely VPU $<$ TPU $<$ Nano $<$  TX2.   Edge accelerators provide performance that is within one-half to 3.5$\times$  that of  x86 server processors. When normalized for power and cost, edge accelerators easily outperform traditional server processors by 10-100$\times$ and become comparable to or better than even server GPUs. 
All edge accelerators exhibit very low power  consumption, ranging from 0.6W to 8W, which is more  than an order of magnitude lower than the server CPU and GPU.
These results indicate that specialized edge architectures are very attractive for edge applications in power or space-constrained settings. Further, they have the potential to replace traditional (``cloudlet-like'') x86 edge servers for deep learning inference workloads.

\section{Split Processing across Application Tiers}

Next, we evaluate the benefits of hardware accelerators for distributed or split processing of edge-based AI workloads.  We consider both model splitting and model compression, which are the two types of split processing that have been proposed previously but have not studied in the context of edge accelerators.

\subsection{Model Splitting}

\begin{figure*}[ht]
    \centering
    \includegraphics[width=\linewidth]{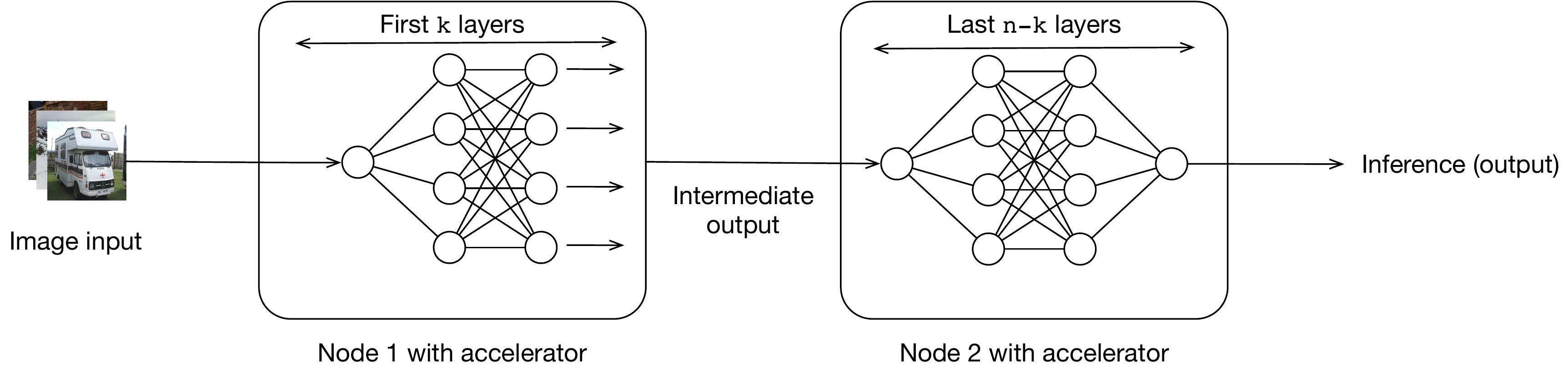}
    \caption{Illustration of Model Splitting of a $n$ layer model, where the first $k$ layers of the model run on node 1 and the intermediate output is sent to node 2 that runs the remaining $(n-k)$ layers.  Nodes 1 and 2 may reside on same tiers (e.g., edge-edge splitting) or on different tiers (e.g., edge-cloud or device-edge splitting)}
    \label{fig:illustration_split}
\end{figure*}

Our first method, {\em model splitting}, which is illustrated in figure \ref{fig:illustration_split} and allows a deep learning model to be split across multiple nodes within or across tiers. In sequential splitting \cite{Kang:2017:NCI:3093315.3037698}, the first $k$ layers of the $n$ layer model run on the first node accelerator and the remaining $n-k$ layers run on the next node or tier. In this case, the inference request is initially sent to the first node and the intermediate output of the $k^{th}$ layer is then sent over the network to the $(k+1)^{st}$ layer running on the second node for subsequent processing.  Model splitting can also be done in parallel, where a portion  of each of the $n$ layers is deployed on the first node, with the remaining portions of each layer deployed on the other node \cite{234801}. In this case, both nodes process the input data in parallel by feeding it through the layers of the model. Model splitting offers two possible benefits. First, in case of sequential splitting, if the output of an intermediate layer is smaller than the input, splitting the model at this layer consumes less network bandwidth  than sending the original input to the higher tier for inference.   Second, model splitting is also useful when the full model does not fit into the memory of a hardware accelerator; in such cases, the  model can be split---sequentially or in parallel---across two or more edge nodes within a tier, enabling  all  processing to be  performed at the edge tier even though no single accelerator can host and run the entire model.

\begin{figure*}[t]
    \centering
    \begin{tabular}{cc}
        \includegraphics[width=2.7in]{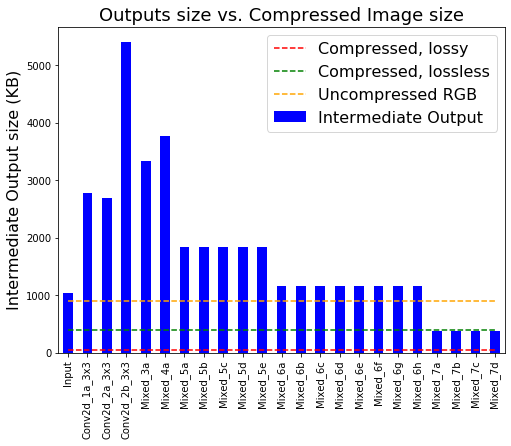}&
    \includegraphics[width=2.7in]{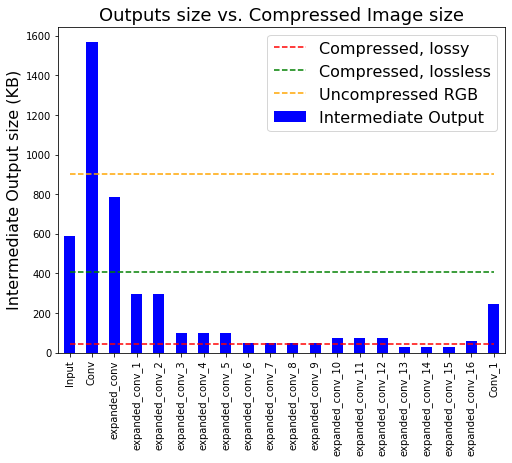} \\
    (a) Inception V4 & (b) MobileNet V2
    \end{tabular}
    \caption{The intermediate output size of various layers of the models for image
    classification.}
    \label{fig:layer_size}
\end{figure*}


 Our first experiment evaluates the benefits of model splitting using sequential splitting for image classification (using our Inception V4 and
MobileNet V2 models). Our experiments were performed by  splitting the model between an Edge TPU and a cloud GPU. For each model, we systematically vary $k$, the layer after which the model is split between the two nodes, and measure the size of the intermediate output  transmitted between layers $k$ and $k+1$. Note that the inference result will always be the same regardless of the chosen $k$, and only the data transmitted between the split models varies with $k$.   We compare this overhead to the non-split model inference where the  entire model runs on a single node, and  the input image data is sent over the network to that node  using (i) uncompressed RGB  format, (ii)  lossless PNG compression
and (iii)  lossy JPEG compression.  

Figure \ref{fig:layer_size}(a) shows the result obtained by splitting the Inception V4 model for 
image classification. As shown, the intermediate output produced
by each layer  varies from layer to layer.  Interestingly, we find that {\em all} 
layers produce an intermediate output that {\em exceeds} the  size of the input data 
when using  using lossless or lossy compression to transmit the input. Only  transmitting the input data in uncompressed RGB format incurs more network overhead.
Thus, splitting at any layer will consume {\em more bandwidth} than sending JPEG or PNG compressed images to a non-split model. 
This result shows that, for Inception V4, there is no benefit from splitting the model
between the edge and the cloud tiers, and it is better to either deploy the full model entirely on  the edge tier and avoid all data transmissions to the cloud, or deploy the model entirely in the cloud by sending compressed inputs to the non-split model.
Further, for Inception V4, the only benefit of  split processing 
is for handling a large memory-footprint model that does not fit into the memory of a single edge accelerator. In this case, we can split the model  across two  (or more) edge
node accelerators to accommodate it and perform distributed inference within the  edge tier using horizontal splitting.

Figure  \ref{fig:layer_size}(b) shows the result obtained by splitting the  MobileNet V2 model for  image classification. We find that the behavior of this model is  different from the previous case. The figure shows that most layers, except for the first two, produce intermediate output that is far below the size of the input data when using  lossless compression.  We find that splitting the model at layer 10 (labelled ``expanded\_conv\_6'') yields nearly 8$\times$ network savings over using lossless compression for a non-split model. 
The  maximum  savings are obtained by splitting at layer 16 (``expanded\_conv\_13'') with nearly an order of magnitude reduction in the used network bandwidth. Splitting
even offers benefits when compared to using lossy JPEG compression, with  layer 16 yielding 30.46\% bandwidth savings. These network bandwidth savings  come with a tradeoff however---the total latency of performing  split inference on two nodes is higher  that performing a single  non-split model inference, as  shown  in 
Table \ref{tab:split}. The table shows that the latency of vertical splitting between the device-edge and edge-cloud tiers as well as horizontal splitting between edge-edge is always higher than the non-split  inference latency (when using VPU, edge TPU and cloud GPU as the accelerators for the device, edge and cloud tiers).
Thus, model splitting involves trading lower network overhead for higher inference latency.

\begin{table}[h]
\centering 
\begin{tabular}{|c|c|c|c|} \hline
Split  &\multicolumn{2}{c|}{Split Latency} & Non-split  \\  \cline{2-3}
between & Node 1 & Node 2  &   Latency \\ \hline
device-edge & 52.19ms  & 4.03ms  & 14.11ms  \\ 
edge-edge & 13.50ms & 4.03ms & 14.11ms \\ 
edge-cloud & 13.05ms & 0.50ms &   1.45ms \\ \hline
\end{tabular}
\caption{Inference Latency for Split vs. Non-split Model. The MobileNet V2 model is assumed  to be optimally split at layer 16. Network latency, which is the same for both, is omitted.}
\label{tab:split}
\end{table}

\begin{figure*}[ht]
    \centering
    \includegraphics[width=\linewidth]{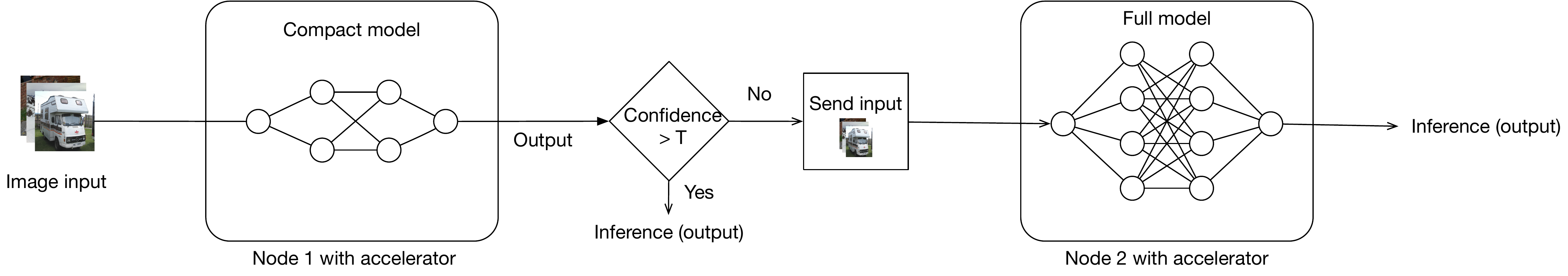}
    \caption{Illustration of model compression. The first node runs a compact model, while the second node runs the full model; the latter model is invoked only when the confidence of the more compact model is below a threshold value}
    \label{fig:illustration_compression}
\end{figure*}

{\noindent \bf Key takeaways:}
Taken together, the results above show that the benefits of model splitting
are highly model dependent. In many cases, significant network savings can be obtained from splitting the model across tiers in an optimal manner, but at the cost of higher  overall inference latency. In other cases, split processing is useful only {\em within} the edge tier when the model does not fit in the memory of a single accelerator, while splitting across tiers is not beneficial from a network standpoint.
Since the overheads and benefits will vary from model to model, adaptive run-time techniques are needed to analyze these overheads and determine whether to split and, if so, an optimal split for each particular model.

\begin{figure*}[t]
    \centering
    \begin{tabular}{ccc}
    \includegraphics[width=2.2in]{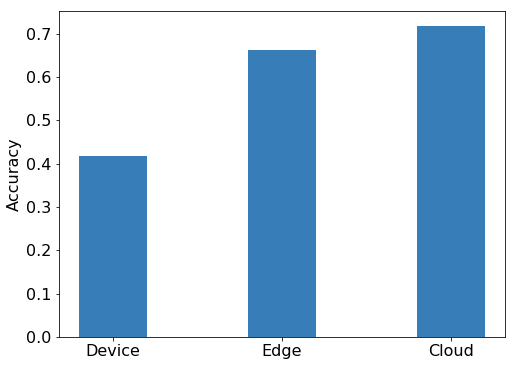} &
    \includegraphics[width=2.2in]{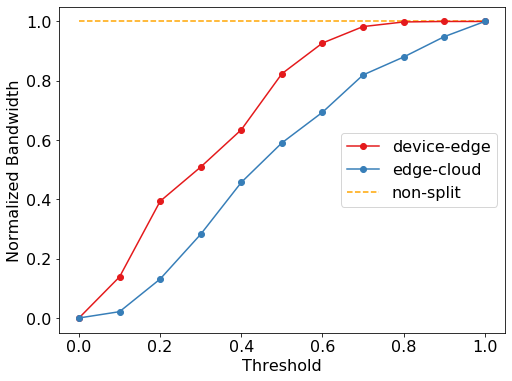} &
    \includegraphics[width=2.2in]{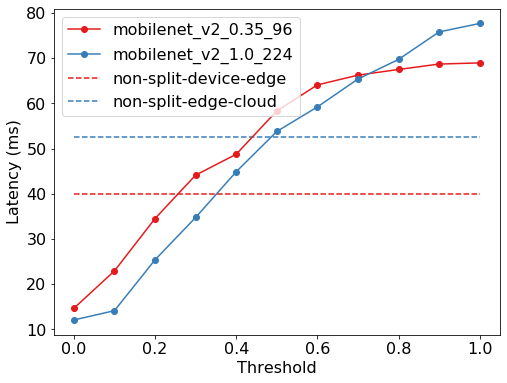} \\
    (a) Accuracy & (b) Bandwidth & (c) Latency  \\
\end{tabular}
    \caption{Accuracy of compressed models (a) and Bandwidth and latency savings for different cut-off confidence thresholds
     under model compression (b,c).}
    \label{fig:compression}
\end{figure*}

\subsection{Model Compression} 

Model compression, as illustrated in figure \ref{fig:illustration_compression}, is an alternative form of split processing that takes a full deep learning model and constructs a smaller compressed version of that model
with a lower memory footprint \cite{Teerapittayanon2017DistributedDN}. The smaller model is deployed for performing inference  on a  lower tier node with less resources, while the full model runs on a more capable  higher-tier node. For example, the small model can be  deployed on the device tier  with a local accelerator, while the larger model runs on an edge node with accelerator capability. Alternatively, the compressed model can be deployed on an edge node with the full model  running  on a cloud server (the difference between these two scenarios is the relative sizes of the  {\em device-edge} and {\em edge-cloud}  models). In either case, inference is first run on the compressed model; since all models produce a probability (confidence) value along with each inference result, the method uses a threshold parameter to determine if the output of the compressed model is of adequate quality, in which case the output is assumed to be final. Otherwise the input data is sent over the network to the full model at the next tier for a second inference.  Such an approach can provide bandwidth and latency savings---if a majority of the inference requests are handled by the compressed model, data need not be sent to the next tier, yielding bandwidth savings, and inference can be  handled locally at lower latencies. The threshold parameter allows for a tradeoff between accuracy, bandwidth, and latency. 

 We now evaluate the efficacy of model compression-based split processing using
hardware accelerators. We consider two scenarios, a \emph{device-edge} case where a very
small footprint model (6.4 MB) runs on the device tier accelerator (emulated using 
a VPU, which is the slowest of our accelerators) along with a larger (13MB) model running on the TX2 edge GPU. We also consider
an \emph{edge-cloud} case where we run a medium footprint (13MB) model on the TX2 edge GPU  and a larger 23 MB model on the cloud GPU. We construct these models of varying size using MobileNet V2, yielding the \texttt{mobilenet\_v2\_0.35\_96} device model, \texttt{mobilenet\_v2\_1.0\_224} edge model and \texttt{mobilenet\_v2\_1.4\_224} cloud model.

Figure \ref{fig:compression}(a) shows the accuracy of the three models on the ImageNet
validation dataset (obtained by comparing the inference results with the ground truth in the dataset).  As can be seen, the smaller the compressed   model, the lower its accuracy. 
Figure \ref{fig:compression}(b)
shows the network bandwidth usage for the device-edge and edge-cloud scenarios
under varying thresholds; recall that the threshold determines the confidence level
under which the input image is transmitted to the next tier for inference  by the larger 
model. A lower threshold implies we are willing to accept predictions with  lower
confidence from the smaller model. As can  be seen, as the threshold increases, a larger 
percentage of inference requests fail to meet the desired confidence using the compressed model and require a second inference from the larger model,
which increases the network bandwidth usage. At a threshold of $0.5$, the device-edge
case yields a 18\% network savings when compared to the non-split scenario; the savings
for the edge-cloud are higher at 41\% since the larger edge model is  able to  handle more inference requests locally than the smaller device model of the device-edge case.
The savings fall to  0.1\% and 11\% 
for a higher threshold of $0.8$ for the device-edge and edge-cloud, respectively, and diminish asymptotically to zero as the confidence threshold approaches 1. 

 Figure \ref{fig:compression}(c) shows the total latency of split processing for different thresholds. The total latency includes the inference latency of the compressed model, the network latency to send data to the larger model if necessary, and the latency of the second inference if the larger model is invoked. For the non-split case, all requests incur network latency to send data to larger model and also include the inference latency of the larger model. In our experiment, the mean edge-device  network latency was around 4ms and the edge-cloud latency to the EC2 cloud server was 47.76ms. In contrast, the inference latency is highest at the device VPU and lowest at the cloud GPU.  
The figure shows that for lower thresholds, split processing offers lower overall latency
since the compressed model is able to produce results of ``adequate'' quality (i.e., above the threshold), which avoids a network hop and a second inference by the larger model.
As the threshold increases, more results need to be sent to the larger model since the compressed model is unable to produce results that meet this higher confidence. This causes the overall latency of split processing to rise due to more requests incurring a network hop and a second inference. 

The figure also shows a cross-over point beyond which split processing  incurs higher overall latency than non-split processing---since the overhead of two inferences is higher than performing a  single inference. We find that the cross-over point occurs at a relatively low threshold of $0.26$ for device-edge and $0.45$
for edge-cloud scenarios.  This implies that when subjected to a {\em random} set of  
inputs (from the Imagenet validation dataset), model compression in not able to outperform 
non-split inference when high confidence output is desired from the smaller model; model compression yields lower latencies only  when we are willing  to accept lower quality results from the compressed model.

\begin{figure}[ht]
    \centering
    \includegraphics[width=\linewidth]{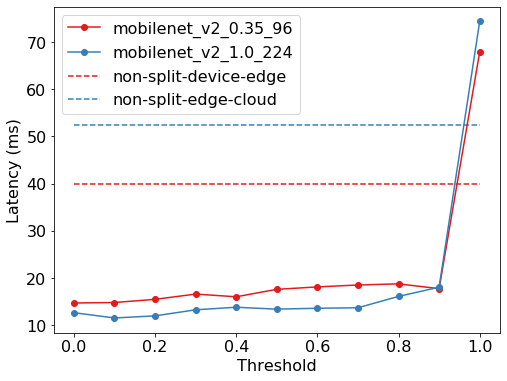}  
    \caption{Latency savings when compressed model produces high confidence results for common case inputs.}
    \label{fig:skewed}
\end{figure}

We next consider a scenario where the inputs are not random but skewed towards the common
case. In this scenario, we assume that the compressed model is well-trained for a small
number of frequently occurring inputs. The larger model is invoked only for less common
inputs for which the compressed model yields less confident and less accurate results.
This is a likely deployment scenario for model compression where the compressed model
is designed to perform well for common case inputs that are frequent, acting as a ``filter'' for such inputs; less common inputs are sent to the larger model, which is capable of handling a much greater range of inputs, for further processing. To evaluate
such a scenario, we construct a skewed input dataset  using the Imagenet validation dataset where common-case inputs (e.g., ``car'')  occur very frequently and all other 
inputs (e.g., all other vehicles) occur infrequently. Figure \ref{fig:skewed} depicts 
the latency of the device-edge and edge-cloud scenario for such inputs. As shown, model
compression yields much lower latency (3$\times$ for device-edge and 4$\times$ for edge-cloud) than non-split inference for a wide range  of threshold values---since it  performs inference well for the common case, and avoids
a second inference for  the majority  of the inputs.  The bandwidth savings (not shown here) are similarly higher than the non-split case for a broad range of threshold values. 

\begin{figure}[ht]
    \centering
    \includegraphics[width=\linewidth]{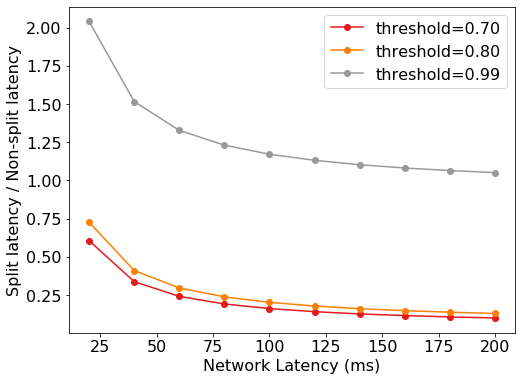} 
    \caption{Inference latency  with varying network latency to cloud servers.}
    \label{fig:cloud-latency}
\end{figure}

Finally, we evaluate the impact of the network latency on these benefits. While the previous experiment used actual network latency to the EC2 cloud server, we evaluate
the benefits of cloud latencies under different emulated cloud latencies. We vary the 
cloud latency from 20ms to 200ms and mesure the  latency of using model
compression relative to the non-split case.  As can be seen in Figure \ref{fig:cloud-latency},
the higher the latency to the cloud server, the greater the benefits of using the compressed model to perform a single local inference. For a theshold of 0.8, 60ms
cloud latency yields around 70.39\% latency reduction and 100ms cloud latency yields a 79.83\% 
lower latency.  The figure also shows that 
higher thresholds yield lower benefits, since it causes more inputs to be sent to  
the larger model. Finally, for very high thresholds such as 0.99,  
split processing is always worse than non-split inference, since it causes the vast majority of the inputs to undergo inference at both the compressed and the larger model.
 

{\noindent \bf Key takeaway:} Unlike model splitting which offers bandwidth savings
by trading off higher latency, model compression can yield \emph{both} bandwidth and latency reduction, but comes with an accuracy tradeoff.
The smaller the compressed model, the lower its ability to perform local inference
with good confidence and accuracy and the lower the bandwidth savings from split processing. Consequently, we find that edge-cloud split
processing yield higher savings than the device-edge case due to the larger
compressed model at the edge. The latency reductions depend significantly 
on the nature of the inputs. 
When optimized for frequent common case inputs, model compression can yield very good latency reduction by handling most  of the frequently occurring inputs locally using the compressed model. The benefits of model compression also depend on the network latency---the higher the latency 
to the cloud, the more valuable is the ability to handle inference locally and avoid
an expensive network hop. Conversely, the closer the cloud servers, the lower are the
benefits of split processing using model compression. 
   

\section{Concurrency and Multi-tenancy}

Our final experiment focuses on concurrency and multi-tenancy considerations for 
specialized edge nodes.  General-purpose nodes are capable of executing  concurrent tenant application due to OS features, such as CPU time sharing  
and address space isolation. 
To understand such benefits for specialized 
edge nodes, we conduct an experiment to quantify the ability of hardware accelerators
to run concurrent  models. To do so, we  load multiple SSD MobileNet V2 models, one for each tenant, onto each of our four edge accelerators.  Each tenant application thread  then invokes its loaded model for inference concurrently with  others. We vary the number of concurrent models and measure the throughput of each device.  

\begin{figure}
     \centering
     \includegraphics[width=\linewidth]{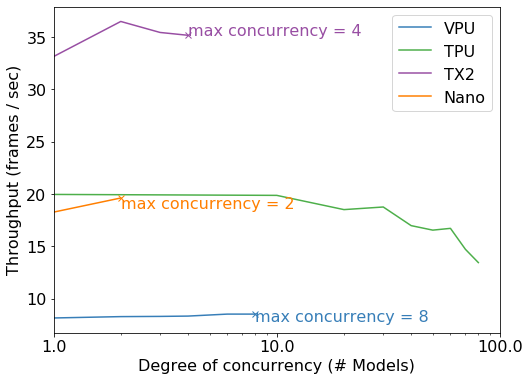}
     \caption{Degree of concurrency supported by various device accelerators.}
     \label{fig:throughput_vs_size}
 \end{figure}

Figure \ref{fig:throughput_vs_size} shows the inference throughput obtained for each hardware accelerator for different degrees of concurrency. The figure shows that all
four edge accelerators are capable of supporting multiple concurrent models and 
provide inference throughput that is comparable to that under
a single tenant scenario. However, the maximum degree of concurrency varies by device.
Typically, the maximum concurrency will depend at least on the device memory size and the 
model size.  For the SSD MobileNet V2  model used in this experiment, the Nvidia 
Nano and TX2 can support a a maximum 
of 2 and 4 concurrent tenants, respectively. Surprisingly, the Intel NCS2 VPU 
can support 8 concurrent models despite being more memory constrained than the 
GPUs.  The Edge TPU has the best concurrency features---it can  arbitrarily scale the number of concurrent models due to its ability to use the host RAM to  store
models that do not fit on the device memory and its use of context switches to swap 
models to and from RAM. When used in conjunction with a Raspberry Pi3 device, we 
are able to scale the number  of concurrent models to 79 before exhausting memory. The figure shows  a slow drop in throughput as we increase the degree
of concurrency due to the increasing  context switch overhead. 

Further analysis revealed that the lower concurrency of the  edge GPUs is due to 
software overheads.  Using the \texttt{nvdia-smi} tools,  we find that each model,
despite being  66MB in size, consumes 1244MB in memory when loaded. This is because
GPUs are designed to be more general accelerators than the VPU and TPU, and its TensorRT
software framework is designed for more general use and therefore more heavyweight (TensorRT libraries alone consume 600MB). In contrast, the VPU and TPU are specifically
designed for deep learning inference and the software framework is heavily
optimized for this use case, thereby imposing low overheads.  

In addition to exploiting host RAM for model swapping, the edge TPU   
also employs model quantization to further reduce memory overheads. Post-training model
quantization \cite{krishnamoorthi2018quantizing} is a technique to  reduce the memory footprint
of the trained model---for example, by quantizing 32bit floating point weights of the model to 8bit precision values. The {\em edgetpu} runtime framework has quantization
turned on by default, enabling it to shrink the size  of each model prior to loading.
The tradeoff though is a possible drop in accuracy of the models due to quantization of the model weights.\footnote{Frameworks such as Tensorflow provide tools to verify that
any such drop in accuracy is within tolerable limits.}   

Finally, we note that none of the devices offer any isolation or security features
for concurrent tenants. Currently a tenant thread can access models belonging to 
other tenants and even overwrite other models. The lack of isolation features implies
that despite supporting concurrent model execution, the devices are not yet suitable
for use in multi-tenant edge clusters or  edge clouds.

\section{Summary and Implications of our Results}
\label{sec:implications}
In this section, we summarize our results and discuss their broader implications.
Our performance experiments revealed that edge accelerators provide comparable or better normalized performance than server CPUs and GPUs, outperform  server CPUs on a raw performance basis, and consume an order of magnitude lower power for inference workloads. Our results imply
that specialized edge clusters can potentially replace x86 edge clusters for such workloads. From a cost standpoint, deploying a large number of edge accelerators is no worse, but ofter better, than deploying a smaller number of more powerful GPUs at the edge.  Specialized edge nodes are especially well suited for power and space constrained settings and open up new possibilities that are infeasible using current  architectures.


Our split processing experiments provided several interesting insights.
We found that model splitting across tiers can offer good bandwidth savings (up to 4$\times$ in our experiments) but this comes at the cost of higher overall latency  due to  running split inference across a network. Even when there are no benefits to be had from splitting models across tiers, split processing {\em within} the  edge tier is still beneficial for running large memory footprint models on constrained edge devices. Since the benefits are highly model dependent, our results point to the need for run-time methods to dynamically determine whether  to split a model and  how to do so optimally.


Unlike model splitting,  model compression can offer {\em both} bandwidth savings and lower inference latency,
but only when a majority of the inference requests can be handled by the compressed
model with  high confidence and accuracy. Highly compressed models or higher confidence
thresholds diminish the benefits of model compression, since they cause a higher fraction
of request to incur a network hop and a second inference.  
Our  results also imply that the latency benefits of model compression will diminish as the latency to the cloud  reduces gradually over time due to the ever increasing number of geographic cloud locations.

Finally, our concurrency experiments show that 
the degree of concurrency depends on the device memory, model size, framework software 
overheads, and system optimizations. Higher device memory does not always translate to
a higher  degree of  concurrency, especially if the run-time  framework is not memory-optimized.
  Conversely, devices with a small amount of memory  can support
a high degree  of concurrency by heavily optimizing the run-time framework and employing
 optimizations, such as model swapping from the host memory and  quantization of the model 
 parameters. However, we find that the lack of isolation and security features between
 the concurrent models is a barrier for their use in  multi-tenant edge cloud environments. 

\section{Related Work}
Recent work on running deep learning applications on the edge falls into three 
categories: (i) cloud-only, (ii) edge-only, and (iii) collaborative edge-cloud.
Cloud-only approaches \cite{gcloud,azure,amazon-ml} allow devices or the edge to offload compute-intensive  inference to the cloud but at the  expense of higher latency. In the context of edge-only approaches, pCAMP has compared various ML frameworks such TensorFlow, Caffe2, and MxNet on various edge devices but without using edge accelerators~\cite{216803}. 
     There has been some initial work on benchmarking edge accelerators.  
  A comparative analyses of keyword spotting audio applications on both cloud and edge devices 
  was performed in \cite{DBLP:journals/corr/abs-1812-01739}; the study
   found that, for keyword spotting audio workloads, edge devices outperform cloud devices on an energy cost per inference basis while maintaining equivalent inference accuracy. 
   The efficacy of running CNNs on low-cost and low-power edge devices for low-power robotics workloads was studied in \cite{Pea2010BenchmarkingOC}. FastDeepIot\cite{Yao-Sensys18} has studied the relationship between neural network structures and execution time to find network configurations that significantly improve execution time and accuracy on mobile and embedded devices. However, unlike us, none of these above efforts have considered split processing within or across tiers.

The notion of collaborative split processing using model splitting or model compression has been studied in several efforts, although not in the context of edge accelerators.
Shadow Puppet\cite{216787} uses edge caching of model results to reduce cloud processing. 
Several techniques to split DNN-based model and partition them between edge and cloud and edge to edge have also been studied.  The Neurosurgeon work uses automatic partitioning of DNN computation between mobile devices and datacenters \cite{Kang:2017:NCI:3093315.3037698}. By offloading smaller intermediate output rather than larger origin inputs, the approach reduces network latency. Further, it run  more resource-intensive portions of the split model  in the cloud, to reduce the mobile energy consumption. In contrast to Neurosurgeon, which uses vertical model splitting between the edge and the cloud, \cite{8493499} proposes a framework that adaptively partitions a CNN-based model horizontally and  distributes concurrent executions of these partitions on tightly resource constrained IoT edge clusters. Similarly, \cite{234801} proposes a dynamic programming-based search algorithm to partition CNNs and run them in parallel using channel and spatial partitioning.  An efficient  method to train distributed deep neural networks (DDNNs) over a distributed computing hierarchy consisting of cloud, edge, and end-devices was proposed in \cite{Teerapittayanon2017DistributedDN}. This method allows the network to make an early decision at the edge when the model confidence is high, thereby reducing communication cost. 
Finally, an incremental approach to partition, offload and incrementally build models on servers was proposed in \cite{Jeong:2018:IIO:3267809.3267828}  
The approach allows the server to execute model inference before the entire model is uploaded, which reduces inference time.  However, accelerators-based splitting has  not been a focus of these efforts.

\section{Conclusions}
 In this paper, we conducted an experimental study to evaluate the benefits and tradeoffs of 
using  specialized edge architectures when compared to traditional edge architectures
 for running edge-based AI applications.
  Our experimental study showed that today's edge accelerators can provide comparable, and in  many cases better, performance, when normalized for power or cost, than edge servers. We found that split processing workloads can yield good bandwidth or latency benefits, but these benefits were highly dependent on how the splitting was done from a model and tier perspective.  
  We found that edge accelerators could support varying degrees of concurrency for 
  deep learning inference,  depending on hardware and software constraints, 
  but lacked isolation mechanisms necessary for cloud-like multi-tenant hosting. Overall, our study found that many open issues still need to be addressed to fully realize the benefits of edge accelerators.

\bibliographystyle{ACM-Reference-Format}
\bibliography{main}


\end{document}